# Relationship between the non linear dynamic behaviour of an oscillating tip–microlever system and the contrast at the atomic scale

J.P. Aimé *, G. Couturier, R. Boisgard, L. Nony

*CPMOH Université Bordeaux 1, 351 Cours de la Libéartion, 33405 Talence, France*

**Abstract**

In this paper, the dynamic behaviour of an oscillating tip–microlever system at the proximity of a surface is discussed. The attractive tip–surface interaction is simply described with a Van der Waals dispersive term and a sphere–plane geometry. We show that the non linear behaviour of the oscillator is able to explain the observed shifts of the resonance frequency as a function of the tip–surface distance without the need of introducing a particular short range force.

*PACS:* 07.79.Lh; 61.16.Ch; 68.35.Bs

*Keywords:* Atomic force microscopy; Mechanics theory; Non linear dynamics of oscillators

## 1. Introduction

Few years ago, the resonant non contact mode has been found as a way to map tip–surface interaction as shifts of the resonance frequency [1,2]. Since then, numerous technical and experimental efforts had been performed showing that measurements of shifts were able to produce images at the atomic scale [3–6]. The 'routinely' achievement of the atomic resolution by several different teams and on different types of surface were thought as a real breakthrough in the field of the scanning probe microscopy. Moreover, the increase of experiments using the resonant non contact mode with a high vibrating amplitude and the systematic use of the intermittent contact (the so called tapping mode) produced numerous stimulating experimental results [7–9]. Therefore, theoretical works dedicated to a non linear analysis of the oscillating behaviour in an attractive and repulsive field were greatly stimulated [10–16].

Before going a step further, it is worth to discuss general ideas at a qualitative level. Leaving aside the technical point that the use of a high amplitude reduces the influence of the fluctuations in frequency [2], a first, counter intuitive, result is the use of a large vibrating amplitude allowing the atomic resolu-tion to be achieved. The most common and shared idea is that to probe an attractive field with an oscillator requires to use a small amplitude in order to keep the force field nearly constant throughout the vibrating amplitude. This comes from the idea that shifts in resonance frequency were uniquely due to the gradient force variations.

---
* Corresponding author. Tel.: +33-5-56-84-89-56; Fax: +33-5-56-84-69-70; E-mail: jpaime@frbdx11.cribx1.u-bordeaux.fr

An image with the atomic resolution based on a repulsive interaction appears quite easily understandable, but getting the same resolution when the attractive Van der Waals forces are involved is much puzzling. The smooth variation of the Van der Waals atom–atom interaction and the finite size of the interacting objects requires a further summation, leading to a tip–sample interaction smoothly varying as a function of the tip–sample distance. For instance, the sphere–plane interaction leads to an attractive force with a $d^{-2}$ power law. The force gradient variations, with a power law $d^{-3}$, should normally not be able to produce features at the sub-atomic scale. This can be straightforwardly deduced within the framework of a linear analysis from which shifts of the resonance frequency as a function of the force gradient is unable to predict the observed shape of the frequency changes as a function of the tip sample distance (see below).

The key point with an AFM is that the recorded information comes from the tip–sample interaction, thus the tip is the collector but that which is measured is the cantilever response. In other words, when the microlever is externally periodically excited, the recorded image is a measure of the oscillating behaviour of the tip–microlever system, which in turn can be very sensitive to change of the strength of the surrounding field either attractive or repulsive.

As a consequence, one can expect to find the experimental conditions for which a small change of the strength of the interacting field induces a spectacular change of the oscillating properties. Such a situation occurs if a non linear dynamic behaviour is observed, such that the oscillator acts as an amplification of a small perturbation. That means that those variations will be able to produce a high contrast on an image even for a small variation of the surface properties.

When the tip–microlever system experiences a purely attractive field, the oscillator exhibits a non linear behaviour [10–12,14]. When the oscillator is set at a drive frequency slightly below the resonance one, a bifurcation from a monostable to a bistable state occurs as soon as the vibrating amplitude is large enough [14]. Such an instability makes the tip–cantilever system very sensitive to any slight change of the tip sample interaction. Indeed, the occurrence of an instability gives an infinite sensitivity with the undesirable effect that the tip can snap the surface. Keeping the amplitude constant, the shifts in resonance frequency show a continuous, monotonous, variation as a function of the tip–surface distance which can be very large when non linear effects take place, and in turn amplifies small changes of the strength of the interaction.

This is the aim of the present work to show that taking into account the non linear behaviour of an oscillator and a Van der Waals like interaction between the tip and the surface could be enough to describe most of the results obtained at the atomic scale. As soon as a high amplitude is used, a non linear behaviour is expected, the larger the amplitude, the sharper is the bifurcation, or more pronounced is the dependence of the shift in frequency as a function of the tip–surface distance. As a consequence, to explain the recorded images a complete understanding of the oscillating behaviour is necessary whenever possible. The paper is organised as follows, in the first part we present the differential equation aiming to describe the resonance behaviour of a non linear oscillator, to do so we discuss the Duffing oscillator that exhibits the main features expected. Then we turn on a more general treatment based on a variational principle from which we do get an exact analytical expression of the harmonic solution. This harmonic solution is discussed and simplified in order to better compare the shift of the resonance frequency as a function of the tip–surface distance given by the linear and the non linear analysis.

## 2. Perturbation theory, approximate solution of the harmonic ansazt

Most of what is given below, can be found in Ref. [17] or Ref. [18], therefore we just recall the basic ideas and underline the most interesting results. The linear equation describing the resonance curve of an oscillator gives the relationship between amplitude and frequency:

$$A(\omega) = \frac{f/m}{\sqrt{(\omega_0^2 - \omega^2)^2 + (2\beta\omega)^2}} \quad (1)$$

with $\beta$ the damping factor giving a quality factor $Q = \omega_0/2\beta$. $f$ and $\omega$ are the external drive force and the drive frequency, respectively. When the tip is far from the surface, the resonance curves are described in Eq. (1).

To take into account a possible non linear behaviour of the oscillator, particularly when attractive or repulsive interaction occur, the Duffing oscillator is a pedagogical and suitable model:

$$\ddot{x} + 2\beta\dot{x} + \omega_0^2 x + \varepsilon\omega_0^2 x^3 = \frac{f}{m}\cos(\omega t) \quad (2)$$

where $\varepsilon < 0$ corresponds to the attractive case and $\varepsilon > 0$ corresponds to the repulsive case. The use of the trial function $A\cos(\omega t - \phi)$ gives shifts in resonance frequency as a function of $\varepsilon$ and of the magnitude of the vibrating amplitude through the relationship [17]:

$$\omega^* = \omega_0\left(1 + \tfrac{3}{4}\varepsilon A^2\right), \quad (3)$$

The shift becomes a function of the amplitude, the larger is the amplitude the larger is the resonance frequency shift (see below). The relationship between the amplitude and the frequency becomes [17]:

$$A(\omega) = \frac{f/m}{\sqrt{\left(\omega_0^2\left(1 + \tfrac{3}{4}\varepsilon A^2\right)^2 - \omega^2\right)^2 + (2\beta\omega)^2}} \quad (4)$$

## 3. Variational method, principle of least action

A variational method can be developed based on the principle of least action [14]. The action $S(x)$ is a functional of the path $x(t)$ and is extremal between two instants.

$$S[x(t)] = \int_{t_a}^{t_b} L(x, \dot{x}, t) \, dt \quad (5)$$

Where $L$ is the Lagrangian of the system. The main aim of the use of the variational principle is to employ a trial function that allows us to perform an analytical treatment. As for the treatment of the Duffing oscillator, we focus on the behaviour of the harmonic solution of the form $x(t) = A(D)\cos(\omega t - \phi(D))$. Using a sphere–plane interaction [19], the Lagrangian is:

$$L = T - U + W = \tfrac{1}{2}m\dot{x}^2 - \left(\tfrac{1}{2}kx^2 + xf\cos(\omega t) - \frac{HR}{6(D-x)}\right) - \gamma x\dot{x} \quad (6)$$

where $H$ is the Hamaker constant, $R$ the radius of the tip, $D$ the distance between the surface and the tip at the equilibrium position at rest and $\gamma$ the damping coefficient. The parameters of the path are $A$ and $\phi$, and the variational principle $\delta S = 0$ becomes a set of two partial differential equations:

$$\frac{\partial S}{\partial A} = 0$$

$$\frac{\partial S}{\partial \phi} = 0$$

after some tedious calculation, we obtain the two coupled non linear equations:

$$\cos\phi = Qa(1 - u^2)\frac{\alpha}{3}\frac{a}{(d^2 - a^2)^{3/2}} \quad (7a)$$

$$\sin\phi = au \quad (7b)$$

where the amplitude, the distance and the frequency are expressed in reduced coordinates $a = A/A_0$ and $d = D/A_0$, $u = \omega/\omega_0$. Where $\omega_0$ and $A_0$ are the frequency and amplitude at the resonance. Finally, we use the dimensionless parameter $\alpha$:

$$\alpha = \frac{HRQ}{m\omega_0^2 A_0^3} \quad (8)$$

$\alpha$ is the ratio between a normalised attractive interaction and the external force. When $\alpha < 1$, a bifurcation from a monostable to a bistable state can be observed, while when $\alpha > 1$, instabilities are not observed and a linear analysis can be used [14]. When a molecule–plane interaction is used, the dimensionless parameter becomes $\alpha = (HR^3Q)/(m\omega_0^2 A_0^5)$ [20].

Solving Eqs. (7a) and (7b) gives the relationships between the amplitude, the distance and the frequency:

$$d\pm = \sqrt{a^2 + \left(\frac{\alpha}{3\left(Q(1-u^2) \pm \sqrt{\frac{1}{a^2} - u^2}\right)}\right)^{2/3}} \quad (9)$$

## 4. Comparison with a simulation and different approaches

Here it is useful to examine an approximation allowing Eq. (9) to be more tractable compared to the one we get with a linear analysis. Consider the case corresponding to:

$$Q \gg \frac{1}{\sqrt{1-u^2}} \quad (10)$$

Eq. (10) means that the relative width of the resonance curve is smaller than that of the relative shift of the resonance frequency. Assuming this condition is fulfilled, when the amplitude is kept constant, Eq. (9) allows a simple relation between the relative shift and the tip–surface distance to be derived. Setting $a = 1$ one gets from Eq. (9):

$$[1-u] \sim 1 - \sqrt{\left(1 - \frac{HR}{3m\omega_0^2 A_0^3(d^2-1)^{3/2}}\right)} \quad (11)$$

Eq. (11) indicates that as soon as condition 10 is fulfilled, the quality factor will not have any influence on the experimental results. In other words, the thinness of the resonance peak does allow a small frequency shift to be recorded, but you might also expect to have the same behaviour even with a smaller quality factor.

For a high amplitude $A_0$ a further simplification can be done which gives an easy comparison with the linear analysis:

$$[1-u] \sim \frac{HR}{6m\omega_0^2 A_0^3(d^2-1)^{3/2}} \quad (12)$$

while a linear analysis gives the frequency shift:

$$[1-u] = \frac{HR}{6m\omega_0^2 A_0^3 d^3} \quad (13)$$

Using the same experimental parameters, Eqs. (12) and (13) predict a completely different behaviour emphasising the influence of the non linear terms in achieving the atomic resolution (Fig. 1).

To go a step further, it is worth comparing the results obtained with the Duffing oscillator and those derived when an oscillator is in an attractive field. To be directly compared, a Taylor's expansion with $x/D$ terms is written without using the reduced coordinate:

$$\ddot{x} + 2\beta\dot{x} + \omega_0^2 x = \frac{f}{m}\cos(\omega t) + \frac{HR}{6mD^2}$$
$$+ \frac{HR}{3mD^3}x + \frac{1}{2}\frac{HR}{mD^4}x^2$$
$$+ \frac{2}{3}\frac{HR}{mD^5}x^3 + \ldots \qquad (14)$$

the term $(HR)/(6mD^2)$ is a static term displacing the position at rest of the cantilever, $(HR)/(3mD^3)$ is the force gradient term, which within the linear analysis gives the frequency shift. Comparison with Eq. (2) gives:

$$\varepsilon = \frac{2}{3}\frac{HR}{m\omega_0^2 D^5} \qquad (15)$$

and with Eq. (3), we have:

$$\omega* = \omega_0\left(1 - \frac{1}{2}\frac{HR}{m\omega_0^2 D^5}A^2\right) \qquad (16)$$

thus, Eq. (16) gives an order of magnitude of the frequency shift expected at a given tip–surface distance.

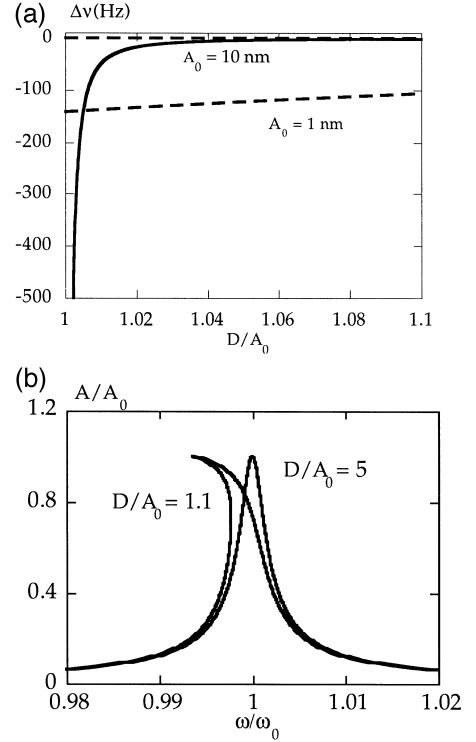

Fig. 1. (a) Comparison between the resonance frequency shift calculated using Eq. (12) (continuous line) and Eq. (13) (dashed line) (see text), with $HR/m\omega_0^2 = 2.81\ 10^{-6}$ J m kg s$^{-2}$. $A_0$ are the resonance amplitudes. (b) Example of distortion of the resonance peak as a function of the tip–sample distance, the resonance peaks are calculated with the help of Eq. (9).

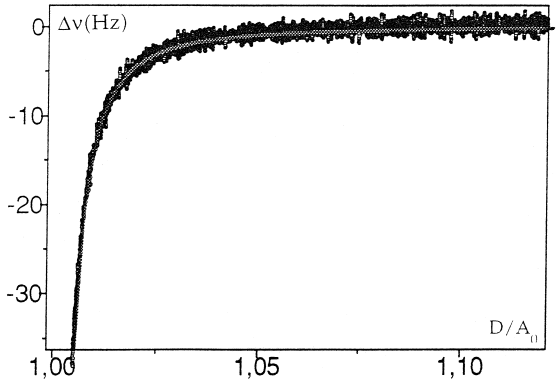

Fig. 2. Fit (continuous line) with Eq. (11) of the numerical results obtained with the simulation of an experiment. The input values in the simulation gives a dimensionless parameter $\alpha = 1.7 \times 10^{-3}$, with a quality factor $Q = 1000$. The refined value with Eq. (12) gives $\alpha = 1.6 \times 10^{-3}$.

To check the validity of the approach, a simulation of the experiments were performed. This was done with the Matlab simulink toolbox. The basis is no more but the one suggested in the Albrecht paper [2]. The blocks of the oscillator and the demodulator needed to measure the frequency shift are simulated. The beginning of the oscillation is provided by a noise generator $b(t)$. The open loop gain is higher than unity, thus an automatic gain control $G(u)$ provides a constant excitation $u(t) = B\cos(\omega t)$. An example of numerical results is reported and compared to the theoretical prediction. The agreement is excellent, the input experimental parameter was the dimensionless parameter $\alpha = 0.00174$ and the refined value obtained from the equation derived from Eq. (9) gives $\alpha = 0.0017$, note also that using the approximate expression given by Eq. (11), satisfactorily describes the overall behaviour with $\alpha = 0.0016$ (Fig. 2).

## 5. Conclusion

The present work aims to describe some of the evolution of the vibrating tip–microlever system. Here we focus essentially on shifts of the resonance frequency that have been shown to be able to record images with the atomic resolution. The theoretical description is able to predict, with a good agreement, the variation of the resonance frequency as a function of the tip–surface distance. It also shows that a linear analysis cannot account for the frequency shift evolution, thus, will be unable to explain large frequency shift when the tip–sample distance is slightly varied. Nevertheless, it remains to transpose this type of work to explain the atomic contrast observed. When a tip is scanned over a surface within the $(X,Y)$ plane, the experiment is different than that recording oscillator properties at a fixed location in the surface and moving up and down along the vertical $z$ axis. As the tip has a finite size and does interact with a more or less extended part of the surface, work has to be performed to explain how small topographic corrugation could be equivalent to what is observed by approaching the surface at a fixed location.

In conclusion, the present paper shows that there is no need to introduce particular short range forces to explain the way the shifts of the resonance frequency is varying as a function of the tip–sample distance.